\documentclass[%
 reprint,
superscriptaddress,
 amsmath,amssymb,
 aps,
]{revtex4-2}

\usepackage{graphicx}% Include figure files
\usepackage{dcolumn}% Align table columns on decimal point
\usepackage{bm}% bold math
\usepackage{physics}
\usepackage{upgreek}
\usepackage{gensymb}
\usepackage{units}
\usepackage[colorlinks=true,linkcolor=blue,urlcolor=blue,citecolor=blue]{hyperref}

\begin{document}

\title{Electrical transport signatures of metallic surface state\\formation in the strongly-correlated insulator FeSb$_2$}

\author{A. G. Eaton}
\author{N. J. M. Popiel}
\affiliation{Cavendish Laboratory, University of Cambridge,\\
 JJ Thomson Avenue, Cambridge, CB3 0HE, United Kingdom}
\author{K.-J. Xu}
\affiliation{Department of Applied Physics and Physics, Stanford University,\\
 Stanford, California 94305, USA}
\author{A. J. Hickey}
\author{H. Liu}
\affiliation{Cavendish Laboratory, University of Cambridge,\\
 JJ Thomson Avenue, Cambridge, CB3 0HE, United Kingdom}
\author{M.~Ciomaga~Hatnean}
\altaffiliation[Present address: ]{Paul Scherrer Institute, Villigen 5232, Switzerland}
\author{G.~Balakrishnan}
\affiliation{Department of Physics, University of Warwick,\\
 Coventry, CV4 7AL, United Kingdom}
 \author{G. F. Lange}
 \author{R.-J. Slager}
 \affiliation{TCM Group, Cavendish Laboratory, University of Cambridge,\\
 JJ Thomson Avenue, Cambridge, CB3 0HE, United Kingdom}
\author{Z.-X. Shen}
\affiliation{Department of Applied Physics and Physics, Stanford University,\\
 Stanford, California 94305, USA}
\affiliation{Stanford Institute for Materials and Energy Sciences,\\
 SLAC National Accelerator Laboratory, Menlo Park, California 94025, USA}
\author{S. E. Sebastian}
\email{suchitra@phy.cam.ac.uk}
\affiliation{Cavendish Laboratory, University of Cambridge,\\
 JJ Thomson Avenue, Cambridge, CB3 0HE, United Kingdom}
 
\date{\today}

\begin{abstract}
\noindent
We present local and nonlocal electrical transport measurements of the correlated insulator FeSb$_2$. By employing wiring configurations that delineate between bulk- and surface-dominated conduction, we reveal the formation of a metallic surface state in FeSb$_2$ for temperatures $\lessapprox 5$~K. This result is corroborated by an angular rotation study of this material's magnetotransport, which also shows signatures of the transition from bulk- to surface-dominated conduction over the same temperature interval as the local/nonlocal transport divergence. Notable similarities with the topological Kondo insulator candidate SmB$_6$ are discussed.

\end{abstract}

\maketitle

Exploring the manifestation of strong electron-electron correlation effects in the bulk of electrically insulating materials has led to several significant advancements in our understanding of quantum matter~\cite{Mott_1949,Anderson-localisation_1958,Kondo1964}. Recently, the exploration of topological insulators -- combining the juxtaposition of a conducting metallic surface state atop a non-correlated insulating bulk -- has also yielded notable results~\cite{Moore464.194,RevModPhys-2010}. The interplay of these two phenomena -- metallic surface states in combination with strong bulk interactions -- may thereby prove to be fertile ground for exploring novel physical behavior.

Iron diantimonide (FeSb$_2$) is a strongly correlated $d$-electron insulator~\cite{Sun-Steglich_AnnPhysik2011} that crystallizes in the orthorhombic P$_{\text{nnm}}$ crystal structure, and exhibits several similarities in common with $f$-electron Kondo insulators~\cite{Tomczak30.183001}. Evidence of electron-electron interactions in this material has been observed in a variety of physical properties, including electrical and thermal transport~\cite{SunPRB2009}, magnetic susceptibility~\cite{Petrovic.72.045103}, optical conductivity~\cite{Herzog82.245205}, specific heat~\cite{Sun_2009}, and x-ray absorption spectroscopy~\cite{Sun_fesb2_spin-states}. At low temperatures $T \approx 10$~K a colossal Seebeck effect has been reported, of $S \approx -45$~mV/K~\cite{Bentien80.17008,Sun_Dalton2010}. While the microscopic origin of this effect is not well understood, it has been conjectured to arise from strongly renormalized carrier masses, the transport of which may be assisted through the phonon-drag effect~\cite{Moore&Phillips,Homes2018,Takahashi7.12732}.

Further effects of the strongly interacting ground state of this material have been uncovered by introducing small doping quantities of Te~\cite{Hu.109.256401}, Co~\cite{kassem2020arxiv}, and Sn~\cite{Bentien.74.205105} which reveal a proximate transition to a heavy-fermion metallic state. A similar doping-induced evolution, from strongly correlated insulator to disordered Fermi liquid, has previously been observed in the related compound iron monosilicide (FeSi)~\cite{DiTusa.78.2831,DiTusa.58.10288,Manyala2008}. FeSi also exhibits several characteristics of Kondo insulator-like strong correlations~\cite{Tomczak30.183001,Schlesinger.71.1748,Mandrus.51.4763}. Recent electrical transport measurements on FeSi have indicated the presence of conductive metallic surface states~\cite{Fang2018} reminiscent of previous findings on samarium hexaboride (SmB$_6$)~\cite{Wolgast-2013,Kim2013-nonlocal}, a 3D topological Kondo insulator candidate~\cite{Dzero104.106408,ann-rev-smb6,Li2020}. While recent angle-resolved photoemission spectroscopy (ARPES) measurements of FeSb$_2$ have indicated the presence of metallic surface states on this material as well~\cite{Ben-ARPES-2020}, complementary electrical transport studies are required in order to probe the interplay between bulk and surface effects.

In this Letter, we present local and nonlocal electrical transport measurements on the correlated insulator FeSb$_2$. Samples were grown by the chemical vapor transport (CVT) method, from starting materials of 99.995\%at Fe and 99.999\%at Sb. Measurements were performed in a Quantum Design Physical Properties Measurement System (PPMS) over a temperature range of 1.7--300~K and up to applied magnetic field strengths of 14~T. Samples were contacted with 25~$\upmu$m diameter gold wire using DuPont 4929N silver paint thinned with 2-butoxyethyl acetate to obtain a suitable consistency with low contact resistance. Some samples also had these connections spot-welded, with no discernible differences between these two preparatory techniques. All contact resistances were measured at room temperature using two-point followed by four-point measurements, indicating combined lead and contact resistances $\sim 1$~$\Omega$ in all cases.

\begin{figure}[t]
\includegraphics[width=1\linewidth]{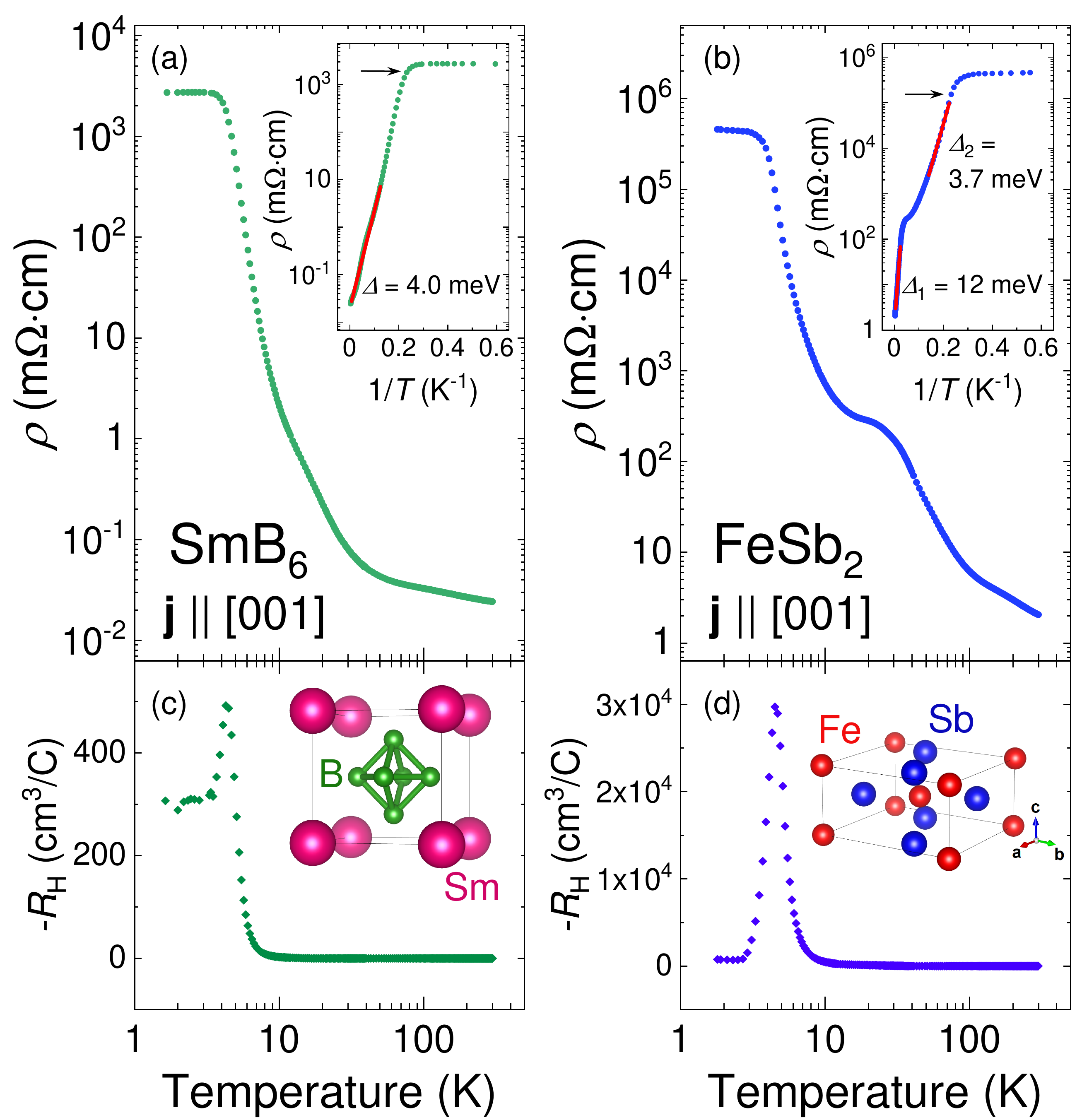}
\caption{\label{fig:compare} Electrical transport anomalies. Resistivity, $\rho$, as a function of temperature for (a) SmB$_6$, and (b) FeSb$_2$, with the Hall coefficient, -$R_H$, plotted on the same logarithmic temperature scale for each in (c) and (d). Both materials show the anomalous formation of a resistance plateau for low temperatures $T \lessapprox 5$~K, which in the case of SmB$_6$ has previously been associated with the development of the low-temperature surface state~\cite{ann-rev-smb6,Li2020}. Upper insets show resistivity against inverse temperature, with arrows indicating the location of the peak in -$R_\text{H}$, which occurs for both materials in close proximity to the resistivity plateau. Red lines show regions fitted to extract activated charge gap energies $\varDelta$. FeSb$_2$ exhibits two gaps, with the lower temperature one, $\varDelta_2$, being of very similar magnitude to the sole gap of SmB$_6$. Lower insets show the respective crystal structures~\cite{VESTA,springer-smb6,springer-fesb2}.}
\end{figure}

Figure~\ref{fig:compare} compares the temperature dependence of electrical transport for SmB$_6$ and FeSb$_2$. In general, the resistivity of a Kondo insulator should increase logarithmically as \textit{T}~$\xrightarrow~0$~K~\cite{Coleman2015book}; however, both SmB$_6$~\cite{Kim2013-nonlocal,Kim2014} and FeSb$_2$ exhibit low-temperature resistance plateaux, a hallmark of the development of a surface state, for $T$~$\lessapprox$~4~K. All sample batches examined in this study exhibited a low temperature resistance plateau (see Supplementary Materials). The formation of this plateau is found to be accompanied by a steep peak in the Hall coefficient, -$R_\text{H}$.

Fitting the FeSb$_2$ resistivity data in Fig.~\ref{fig:compare}b to a doubly gapped Arrhenius activation model, of the form $\rho \propto \exp(\nicefrac{\varDelta}{k_{\text{B}}T})$, yields a high temperature indirect bandgap, $\varDelta_1$, of $\approx$~12~meV for $T >$~30~K, comparable with prior reports~\cite{Tomczak30.183001}. However, at lower temperatures another gap -- $\varDelta_2 \approx$~3.7~meV -- opens up, of comparable magnitude to the 4.0~meV gap in SmB$_6$ attributed to Kondo-induced band hybridisation~\cite{Li2020}. Very similar observations, of low-temperature resistivity saturation concomitant with a dramatic reduction in $R_\text{H}$, have previously been attributed with topologically protected surface state formation in the non-correlated topological insulators ${\text{Bi}}_{2}{\text{Te}}_{2}\text{Se}$~\cite{Ando-Bi-TI} and Bi$_{1.08}$Sn$_{0.02}$Sb$_{0.9}$Te$_{2}$S~\cite{Natcomms-Bi-TI}. The notable similarities between these materials and FeSb$_2$ are strongly suggestive of similar microscopic mechanisms at play in this system.

\begin{figure*}[t]
\includegraphics[width=1\linewidth]{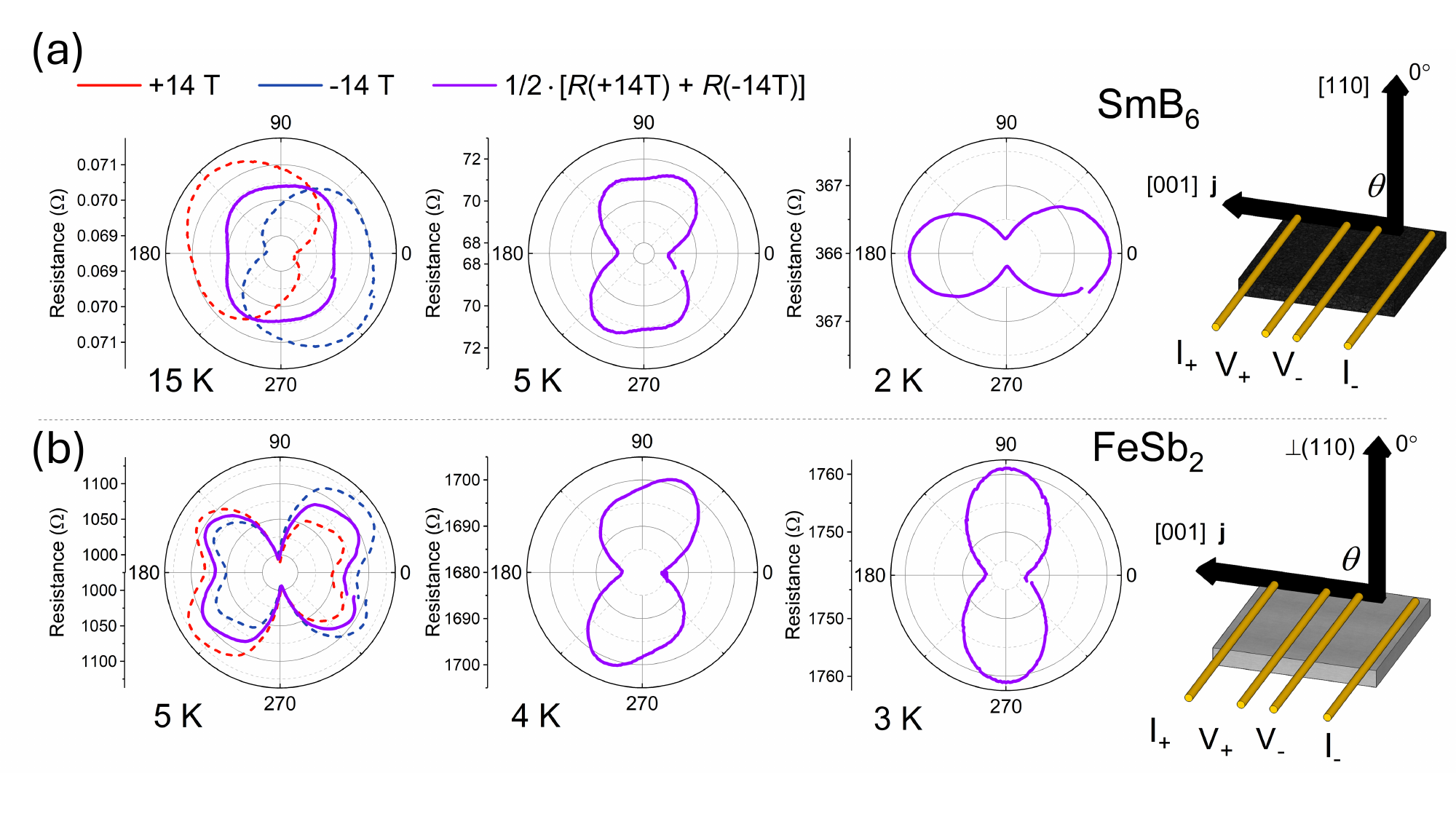}
\caption{\label{fig:rotation} Magnetotransport evolution from bulk- to surface-dominated conduction on cooling. All traces performed with a constant dc current of 25 $\mu$A for $T \leq 5$~K and 100 $\mu$A for $T > 5$ K. (a) Magnetoresistance of SmB$_6$ at successive temperatures as a function of rotation angle, $\theta$. In all panels, the right hand side schematic gives the applied current direction, \textbf{j}, and rotational plane, while purple curves in each plot give symmetrised resistance for applied magnetic fields of $\pm$~14~T, to exclude any transverse contributions. At 15~K, away from the resistivity plateau, SmB$_6$ exhibits $C_{4}$ rotational symmetry, whereas at 2~K, on the plateau, this has reduced to $C_{2}$ symmetry. (b), Similar temperature progression on an FeSb$_2$ crystal as for SmB$_6$ in panel (a), clearly exhibiting a very similar peanut-shaped trace at 2~K.}
\end{figure*}

To further investigate the potential surface state origin of the low-temperature resistivity plateau of FeSb$_2$, a rotational magnetotransport study was performed. Figure~\ref{fig:rotation} shows the angular-dependent magnetoresistance (ADMR) of two FeSb$_2$ samples and one SmB$_6$ sample as a function of rotation angle $\theta$, at various temperatures in applied magnetic fields of $\pm$~14~T. Samples with large surface to volume ratios were chosen, with $\theta = 0 \degree$ corresponding to field applied normal to the two largest (parallel) faces, the top one of which was contacted with gold wires for standard four-point electrical transport measurements as indicated in the schematics in Fig.~\ref{fig:rotation}. For $T \geq$~15~K, well away from the plateau, both materials' longitudinal magnetoresistances exhibit almost isotropic rotational symmetry, as expected for simple bulk conduction. However, on cooling to $T \leq$~3~K, within the plateau region, the rotational magnetoconductance profiles of both SmB$_6$ and FeSb$_2$ undergo a pronounced evolution. In Fig.~\ref{fig:rotation}A at 5~K the studied SmB$_6$ sample exhibits nodes (antinodes) in the magnitude of the magnetoresistance at 0$\degree$ and 180$\degree$ (90$\degree$ and 270$\degree$); however, at 2~K this has swapped to give nodes (antinodes) at 90$\degree$ and 270$\degree$ (0$\degree$ and 180$\degree$). Comparatively, the FeSb$_2$ sample in Fig.~\ref{fig:rotation}b evolves similarly -- with a stark difference between the 5~K and 4~K temperature points, whereby the nodes at 90$\degree$ and 270$\degree$ at 5~K have shifted to 0$\degree$ and 180$\degree$ at 4~K.

Interestingly, this evolution goes through a drastic change around the temperature of the Hall effect peak (Fig.~\ref{fig:rotation}), with a convolution of 3D and 2D transport signatures observed in the `butterfly' shape of the 5~K trace in Fig.~\ref{fig:rotation}b -- similar behaviour has previously been reported in studies of Weyl semi-metals~\cite{butterfly_ZrSiS}. At the lowest temperatures, a distinct peanut-shaped trace is observed in the rotational magnetoresistance of FeSb$_2$, in marked similarity with prior SmB$_6$ reports~\cite{chen-flutterbies,Japan-flutterbies}.

This symmetry evolution fits well within the surface state model, whereby at the lowest temperatures electrical conduction is dominated by these surface states, as their conductivity is much higher than that of the insulating bulk. Hence, for the samples measured here with two large, dominant flat faces (as depicted in Fig.~\ref{fig:rotation}), we can conclude that a model comprising the presence of surface-dominated conduction for $T \lessapprox 6$~K, and bulk-dominated conduction at elevated temperatures, very well captures the symmetry evolution observed in the ADMR data as a function of temperature for FeSb$_2$ (and for SmB$_6$).

\begin{figure}[t]
\includegraphics[width=1\linewidth]{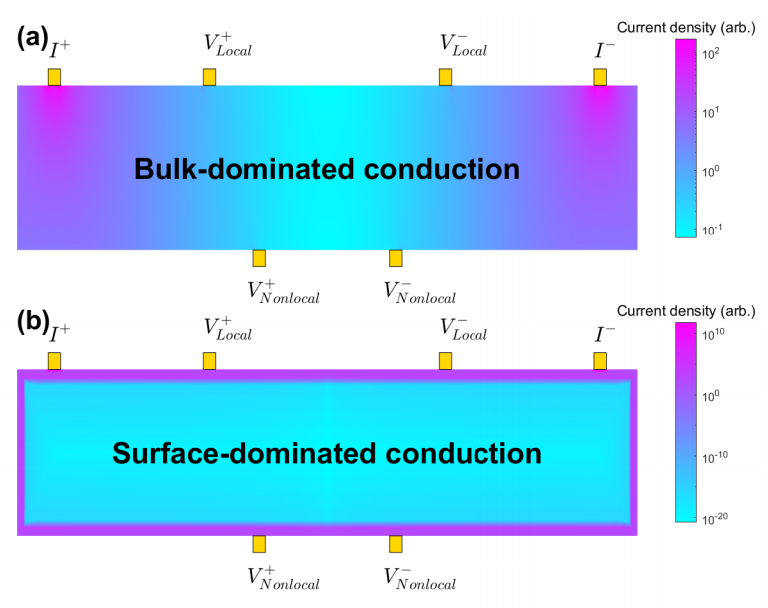}
\caption{\label{fig:FEA} Bulk- and surface-dominated conduction. Finite element simulations of the current density distributions for electrical transport through (a), bulk-dominated channels, and (b), for surface-dominated conduction.}
\end{figure}

\begin{figure}[t]
\includegraphics[width=0.9\linewidth]{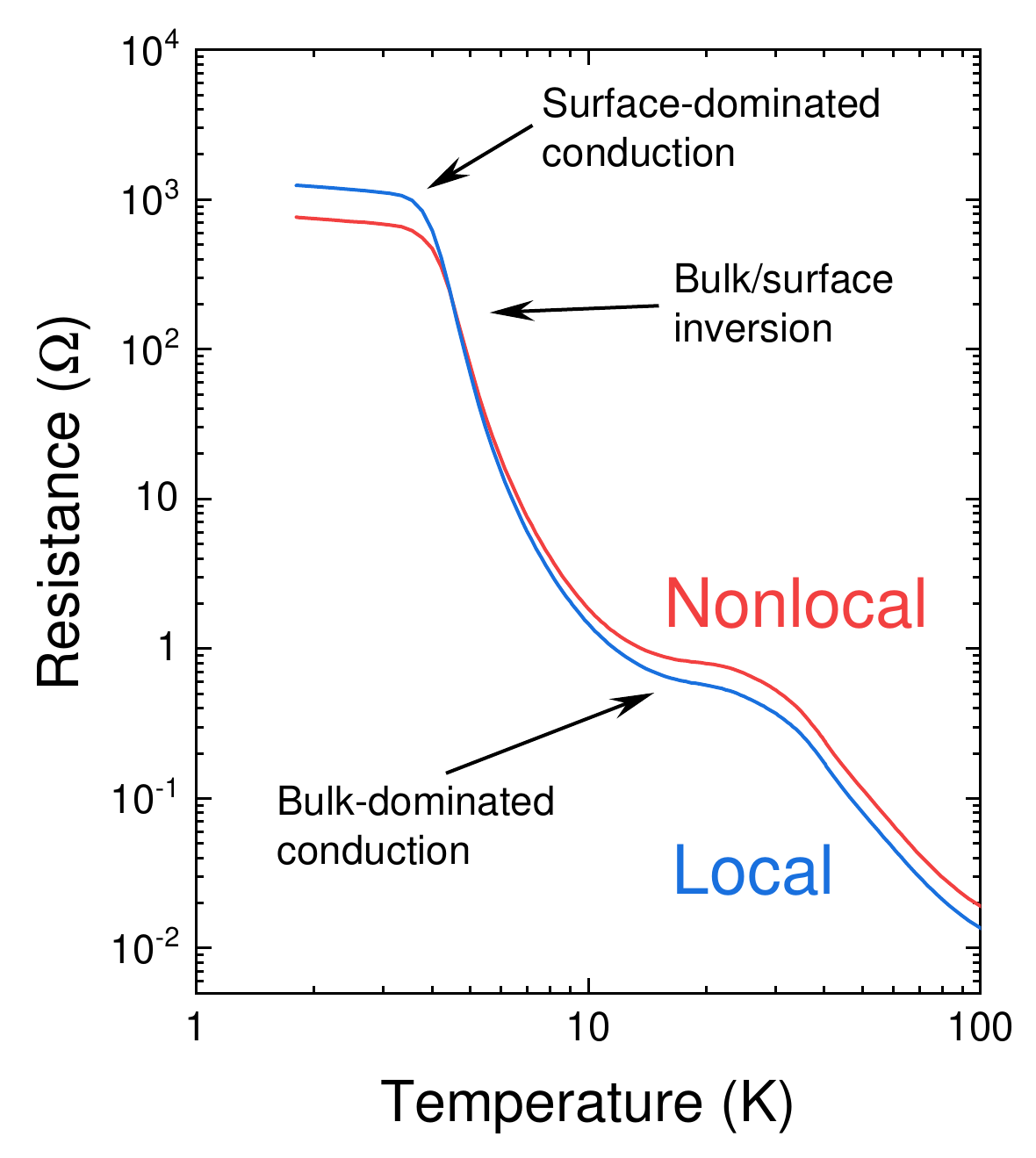}
\caption{\label{fig:nonlocal} Local and nonlocal 4-wire electrical transport measurements performed on an FeSb$_2$ single crystal. Current and voltage leads were attached as per the configurations indicated in Fig. 3. At high temperatures the nonlocal resistance is greater than the local resistance, as expected for simple bulk conduction. However, at low temperatures the converse is true, with the local resistance exceeding that of the nonlocal resistance. This inversion is characteristic of the formation of a conductive surface state.}
\end{figure}

To further probe the character of low-temperature electrical conduction in FeSb$_2$, a nonlocal resistance study was undertaken. Previously, an inversion between local and nonlocal electrical transport measurements upon cooling to the resistivity plateau was taken as a key piece of evidence in support of metallic surface state formation in SmB$_6$~\cite{Li2020,Wolgast-2013}. In the general case of electrical conduction through a section of bulk 3D conductor, electrical transport is expected to take place predominantly through the bulk of the material, as indicated by the current density distribution depicted in Fig.~\ref{fig:FEA}a. However, for a sample cooled to $T$~$\lessapprox$~4~K, the surface state model predicts that band hybridisation will cause these bulk conduction channels to become strongly gapped out as the surface state forms. This results in electrical transport being constrained almost entirely to the metallic surface of a sample (Fig.~\ref{fig:FEA}b). By performing four-wire resistance measurements of local and non-local configurations, as shown by Figure~\ref{fig:FEA}, the applicability of this bulk/surface model to FeSb$_2$ may be investigated, and compared to the excellent correspondence previously found for such measurements in SmB$_6$~\cite{Li2020,Wolgast-2013}.

Fig.~\ref{fig:nonlocal} shows the measured electrical resistance of an FeSb$_2$ crystal for local (blue curve) and nonlocal (red curve) wiring configurations. At elevated temperatures ($T \gg$~4~K) the resistance of the non-local transport channel is greater than the local channel, as expected by simple geometric resistivity arguments for bulk conduction. However, for $T$~$\lessapprox$~5~K the reverse is true, with $R_{\text{local}} > R_{\text{non-local}}$. This crossover occurs just before the resistivity plateau, over the same temperature range as the Hall effect peak in Fig.~\ref{fig:compare}. Such a remarkable observation would be difficult to reconcile within a bulk impurity-band picture, or even with an elaborate low-temperature highly-anisotropic conductivity model. However, these observations may be easily explained by the surface state picture as applied to very similar observations in SmB$_6$~\cite{Li2020,Wolgast-2013}, and thus provide compelling evidence that the same phenomenon is occurring in FeSb$_2$. We note that these measurements are in strong agreement with a recent corbino disc study~\cite{eo_extraordinary_2023}.

Furthermore, the fact that the local/non-local resistance inversion should occur over the same small temperature range (approximately 4 to 6 K) as the peak in the Hall coefficient, the opening of the resistance plateau, and the magnetoresistive symmetry evolution, each in concurrence with previous reports on SmB$_6$, gives strong confidence that these observations all point to the same correlation-driven surface state mechanism, previously investigated in the case of SmB$_6$, also being present in FeSb$_2$.

To investigate the possibility of non-trivial topology manifesting the observed surface states in FeSb$_2$, we performed DFT calculations to evaluate the crystalline symmetry (see Supplemental Materials). However, from these calculations we find that $z_4 = 0$, indicating that the bandstructure of FeSb$_2$ is topologically trivial. We note that our DFT calculations are limited to assuming a nonmagnetic groundstate as previously reported~\cite{Petrovic2003} -- however, the presence of localized Fe 3d states may necessitate a more detailed calculation, fully accounting for the strong electronic correlations, to fully model the true bandstructure. Regardless of the possibility of a topologically trivial or nontrivial origin, the metallic surface state in FeSb$_2$ represents a fascinating platform on which to explore the interplay between strongly correlated physics and reduced dimensionality.

It should be remarked that a clear-cut consensus on SmB$_6$ being a topological Kondo insulator is yet to reach unanimous agreement throughout the community; although, recently this appears to be nearing~\cite{Li2020}. That these measurements on FeSb$_2$ presented here yield a plethora of striking similarities with SmB$_6$ opens a number of pertinent questions, especially regarding the role of $f$-electron physics -- previously thought to be responsible for driving the surface state mechanism in SmB$_6$, but obviously not present in the case of FeSb$_2$.

Studies on SmB$_6$ into what physical features are fundamental properties required for the manifestation of the (putatively topological) surface state, and what features are not relevant, have been severely hampered by drastic differences between sample preparation methods~\cite{Li2020}. Samples grown in aluminum flux are reasonably easy to produce, however the presence of aluminum inclusions, inhomogeneous growth rates of different crystalline facets, and contamination from the melt container, have led to notable inconsistencies in observed properties between different research groups~\cite{Phelan2016}. The presence of small quantities of rare-earth impurities, such as gadolinium, can have a pronounced effect on the physical properties~\cite{Fuhrman2018} at the low temperatures where the surface state forms, and the purification of elemental lanthanides from extraneous impurities of other rare-earth contaminants is notoriously difficult.

Higher purity SmB$_6$ samples may be obtained via the floating zone technique~\cite{Monica-SmB6-growth}, as evidenced by these crystals' superior inverse residual resistance ratios, magnetisation, quantum oscillation and thermal transport properties~\cite{Mate-aluminium}. Samples grown by this method have also exhibited some of the most compelling ARPES evidence for a topologically non-trivial nature of the surface state~\cite{Japan-SmB6-ARPES-111-FZ}. However, this method is considerably more difficult to perform than flux growth, with the high melting point of SmB$_6$ requiring the use of mirror furnaces to reach temperatures approaching $\sim$~3,000~K~\cite{Monica-SmB6-growth}. Furthermore, the attainment of such high temperatures increases the likelihood of introducing samarium vacancies, which have also been reported to be a significant variable factor in the correlated electronic properties of this material~\cite{eoFGFZ}. These variations in physical properties between differing growth batches and sample preparation methodologies of SmB$_6$ has hampered the identification of what constituent properties are prerequisite to the development of the correlation-induced surface state.

FeSb$_2$, by contrast, has a much lower melting point of just over 1,000~K. This makes the acquisition of high-purity samples much more accessible through use of the CVT method, such as for the crystals investigated in this study. The CVT method, like the floating zone method, is immune to extraneous substitutive impurities from a crucible origin, or from flux inclusions; importantly, the considerably lower temperatures involved significantly minimize the probability of introducing vacancies, dislocations, or other sources of crystalline disorder that higher temperature procedures suffer from. Hence, obtaining excellent purity FeSb$_2$ crystals is much more realizable for a broader number of experimental research groups than is the attainment of similarly high quality SmB$_6$.

Therefore, given the striking similitude in low-temperature surface state properties between FeSb$_2$ and SmB$_6$, an exciting possibility lies in the identification of FeSb$_2$ as a model $d$-electron analogue of SmB$_6$, with extremely similar low-temperature behaviour, but with chemical properties that make the attainment of pristine high-purity specimens significantly easier to realize.

In conclusion, we have experimentally observed evidence of low-temperature surface-dominated conduction in the strongly correlated $d$-electron insulator FeSb$_2$ through measurements of local and nonlocal electrical transport. This interpretation is corroborated by an angular rotation study of this material's magnetotransport, which shows an associated evolution from bulk- to surface-dominated symmetry on cooling to temperatures $\lessapprox 5$~K. Complementary measurements, such as scanning tunneling microscopy and spin-polarised ARPES, would be instructive in resolving any possible topological character of these surface states, and may reveal further manifestations of exotic correlation-driven effects previously observed in similar systems.

\clearpage

\bibliography{FeSb2}

\end{document}